\documentclass[final,3p,times]{elsarticle}
\biboptions{sort&compress}

\usepackage{amsmath}
\usepackage{amssymb}
\usepackage{bm}
\usepackage{cancel}
\usepackage{feynmf}
\usepackage{graphicx}
\usepackage[pdftex,colorlinks,hyperindex,plainpages=false,bookmarksopen,bookmarksnumbered]{hyperref}

\allowdisplaybreaks
\let\eps=\epsilon
\let\de=\partial
\newcommand\gE{\gamma_\text{E}}
\newcommand\vek[1]{\bm{#1}}
\newcommand\gr[1]{\text{#1}}
\newcommand\imag{\text{i}}
\newcommand\dd{\mathop{\text d}\nolimits}
\newcommand\sumint[1]{\int\kern-1.4em\sum\nolimits_{#1}} 
\newcommand\mii{M_\text{I}}
\newcommand\miii{M_\text{II}}
\newcommand\F[1]{z_{#1}}
\newcommand\Iup[1]{{#1}^\text{I}}
\newcommand\IIup[1]{{#1}^\text{II}}
\newcommand\K[1]{\mathcal{K}_{#1}}
\newcommand\X{\mathcal{X}}

\begin{document}
\begin{fmffile}{fig}

\begin{frontmatter}

\title{Two-loop free energy of three-dimensional antiferromagnets\\
in external magnetic and staggered fields}

\author[tb]{Tom\'a\v{s} Brauner\corref{cor}}
\ead{tomas.brauner@uis.no}
\cortext[cor]{Corresponding author}
\address[tb]{Faculty of Science and Technology, University of Stavanger, 4036 Stavanger, Norway}

\author[cph]{Christoph P.~Hofmann}
\address[cph]{Facultad de Ciencias, Universidad de Colima, Colima C.P.~28045, Mexico}

\begin{abstract}
Using a model-independent low-energy effective field theory, we calculate the free energy of three-dimensional antiferromagnets in a combination of mutually perpendicular external magnetic and staggered fields at the next-to-next-to-leading, two-loop order. Renormalization is carried out analytically, and the renormalization group invariance of the result is checked explicitly. The free energy is thus expressed solely in terms of temperature, the external fields, and a set of low-energy coupling constants, to be determined by experiment or by matching to the microscopic model of a given concrete material.
\end{abstract}


\end{frontmatter}


\section{Introduction}
\label{sec:intro}

The low-energy and low-temperature properties of antiferromagnetic insulators are dominated by their soft excitations: the spin waves (magnons). The analysis of spin systems using a theory of these collective excitations and their interactions has a long history (see Refs.~\cite{Dyson:1956zza,Oguchi:1960zz,Akhiezer1,Akhiezer2,Keffer,Halperin:1969zz} for some of the original works and Ref.~\cite{PhillipsRosenberg} for an early review). However, only relatively recently has one started to approach the problem using the full power of the model-independent effective field theory (EFT) formalism~\cite{Hasenfratz:1993vf,Leutwyler:1993gf,Hofmann:1997qm,Roman:1999ro,Hofmann:1998pp,Roman:canted,Hofmann:2011dm}. In this paper, we consider a case of particular interest: antiferromagnets in an external magnetic field. We carry out, for the first time, an EFT analysis of this system at the next-to-next-to-leading order of the derivative expansion, that is at two loops. We focus on three-dimensional antiferromagnets; the technically simpler case of two-dimensional antiferromagnets in external magnetic and staggered fields was addressed in the preceding paper~\cite{Hofmann:2017mrt}. Just as therein, we also assume the presence of an external staggered field, perpendicular to the magnetic field; this plays the role of a symmetry-breaking perturbation that gives both magnons a nonzero gap.

The paper is organized as follows. In Section~\ref{sec:EFT} we review the basics of the low-energy EFT for antiferromagnets and discuss the magnon spectrum in external magnetic and staggered fields. Some auxiliary details regarding the construction of the effective Lagrangian are deferred to~\ref{app:lagrangian}. In Section~\ref{sec:1loop} we describe the basic setup for the calculation of the free energy using the imaginary-time formalism. In order to introduce our notation and to explain the methodology in as simple a setting as possible, we first show how to determine the free energy at the leading (LO) and next-to-leading (NLO) order of the derivative expansion, which amounts to evaluating one-loop diagrams and the necessary counterterms. The full next-to-next-to-leading-order (NNLO) calculation, including two-loop contributions to the free energy, is postponed to Section~\ref{sec:2loop}. Finally, in Section~\ref{sec:summary} we summarize and conclude.

While the physical implications of the achieved result for the two-loop free energy of three-dimensional antiferromagnets are discussed in a companion paper~\cite{companion}, here we focus on the methodology and the details of the computation, which include a number of novel aspects in their own right. This applies in particular to the calculation of the sunset diagram at nonzero temperature and with two different masses, detailed in Section~\ref{subsec:sunset} and~\ref{app:sunset}, but also to the detailed justification of the implementation of both the magnetic and the staggered field in the effective Lagrangian, given in~\ref{app:lagrangian}.


\section{Low-energy effective theory of antiferromagnets}
\label{sec:EFT}

In the absence of spin-orbit coupling, antiferromagnets possess an \emph{internal} global $\gr{SO}(3)$ symmetry corresponding to continuous spin rotations. The spin alignment in the ground state at zero temperature breaks this symmetry down to the $\gr{SO}(2)$ subgroup.\footnote{This is an exact statement that does not rely on approximating the true ground state with the semi-classical N\'eel state.} The spontaneous breaking of the spin rotation symmetry gives rise to two Nambu--Goldstone bosons---the magnons---which, in absence of other gapless modes in the spectrum, dominate the low-energy physics of antiferromagnets. The dynamics of magnons is described by a low-energy EFT whose form is fully dictated by symmetry except for a few low-energy coupling constants (LECs), to be determined by experiment or by matching to an underlying microscopic theory~\cite{Weinberg:1978kz}. The EFT is therefore model-independent in the sense that it correctly reproduces the predictions of \emph{any} microscopic model with the same symmetry; all dependence on the microscopic dynamics is absorbed in the values of the LECs.

A precise algorithm for constructing the effective Lagrangian, valid for an arbitrary pattern of breaking of internal symmetry, has been known for nearly five decades~\cite{Callan:1969sn}. Here we will follow the more conventional setup in which the magnons are represented by a unit vector field $\vec U(x)$, in line with the fact that the coset space of broken symmetry, $\gr{SO(3)/SO(2)}$, is equivalent to a sphere, $S^2$; the correspondence of this picture with the general setup of Ref.~\cite{Callan:1969sn} is clarified in~\ref{app:lagrangian} and in Ref.~\cite{Andersen:2014ywa}. Due to the linear dispersion relation of antiferromagnetic magnons in absence of symmetry-breaking perturbations such as external fields, the low-energy EFT possesses a pseudo-Lorentz invariance, only differing from the true Lorentz invariance of elementary particle physics by a different value of the fundamental speed, here represented by the phase velocity of magnons. We will use this emergent Lorentz invariance to constrain the form of the effective Lagrangian.


\subsection{Effective Lagrangian}
\label{subsec:Lagrangian}

The effective Lagrangian is constructed by imposing the continuous space and time translation, Lorentz and internal $\gr{SO}(3)$ invariance. The basic building blocks for the construction of the Lagrangian are:
\begin{itemize}
\itemsep0pt
\item The unit vector $\vec U(x)$, transforming as a scalar under Lorentz transformations and as a vector under $\gr{SO}(3)$.
\item Its covariant derivative $D_\mu\vec U(x)$, where
\begin{equation}
D_\mu\vec U\equiv\de_\mu\vec U+\delta_{\mu0}\vec H\times\vec U,
\end{equation}
and $\vec H(x)$ is the external magnetic field. It includes by definition the magnetic moment for the microscopic spin degrees of freedom.
\item Possibly higher-order covariant derivatives of $\vec U(x)$.
\item The staggered field $\vec s(x)$, transforming as a scalar under Lorentz transformations and as a vector under $\gr{SO}(3)$.
\end{itemize}
The Lagrangian is organized according to a derivative expansion, wherein (covariant) derivatives count as order one and the staggered field $\vec s(x)$ counts as order two. This is completely equivalent to the chiral perturbation theory of strong nuclear interactions, where $\vec s(x)$ corresponds to the quark masses~\cite{Gasser:1983yg,Gasser:1984gg}.

Thanks to the assumed Lorentz invariance, only terms with even orders in the derivative expansion exist in the effective Lagrangian in three spatial dimensions. The leading, second-order Lagrangian takes the conventional form
\begin{equation}
\mathcal L_\text{eff}^{(2)}=\frac12F^2D_\mu\vec U\cdot D^\mu\vec U+\vec s\cdot\vec U.
\label{Lagrangian_LO}
\end{equation}
The effective coupling $F$ equals the square root of the spin stiffness, and corresponds to the pion decay constant in the chiral perturbation theory. The staggered field $\vec s(x)$ itself plays the role of the effective coupling in the second term. Note that the LO Lagrangian~\eqref{Lagrangian_LO} possesses an emergent, or accidental, parity symmetry. At the NLO, the underlying crystal lattice may induce perturbations that violate both the continuous Lorentz invariance and the discrete parity symmetry of the LO theory~\cite{Roman:1999ro}. However, these do not affect the renormalization problem, discussed in this paper, and can thus be added to the EFT afterwards. We shall therefore impose both symmetries at the NLO level as well.

With the above limitation, the next-to-leading, fourth-order Lagrangian contains, in presence of a gauge field for the $\gr{SO}(3)$ symmetry, the following independent operators,
\begin{equation}
\begin{gathered}
(D_\mu\vec U\cdot D^\mu\vec U)^2,\quad
(D_\mu\vec U\cdot D_\nu\vec U)^2,\quad
(\vec s\cdot\vec U)(D_\mu\vec U)^2,\quad
(\vec s\cdot\vec U)^2,\quad
\vec s^2,\\
\vec F_{\mu\nu}\cdot\vec F^{\mu\nu},\quad
(\vec F_{\mu\nu}\cdot\vec U)^2,\quad
\vec F_{\mu\nu}\cdot(D^\mu\vec U\times D^\nu\vec U),\quad
D_\mu D^\mu\vec U\cdot D_\nu D^\nu\vec U,\quad
(\vec s\times\vec U)\cdot(D_\mu D^\mu\vec U),
\end{gathered}
\end{equation}
where $\vec F_{\mu\nu}\equiv\de_\mu\vec A_\nu-\de_\nu\vec A_\mu+\vec A_\mu\times\vec A_\nu$ is the field-strength tensor for the $\gr{SO}(3)$ gauge field. However, only the operators on the first line are relevant for us. First, in our case only the temporal component of the $\gr{SO}(3)$ gauge field is nonzero and equal to $\vec H$, and thus $\vec F_{\mu\nu}=0$. Second, the last two operators on our list can be eliminated in favor of the others by using the equation of motion following from the LO Lagrangian~\eqref{Lagrangian_LO}. All in all, the NLO Lagrangian takes the form
\begin{equation}
\mathcal L^{(4)}_\text{eff}=e_1(D_\mu\vec U\cdot D^\mu\vec U)^2+e_2(D_\mu\vec U\cdot D_\nu\vec U)^2+\frac{k_1}{F^2}(\vec s\cdot\vec U)(D_\mu\vec U)^2+\frac{k_2}{F^4}(\vec s\cdot\vec U)^2+\frac{k_3}{F^4}\vec s^2,
\label{Lagrangian_NLO}
\end{equation}
where $e_{1,2}$ and $k_{1,2,3}$ are the LECs; the powers of $F$ were inserted in order to make these couplings dimensionless in three spatial dimensions.


\subsection{Ground state and excitation spectrum}
\label{subsec:spectrum}

The ground state of the antiferromagnet in presence of \emph{uniform} external fields $\vec H$ and $\vec s$ is obtained by maximizing the static part of the effective Lagrangian,
\begin{equation}
\mathcal L^{(2)}_\text{eff,stat}=\frac12F^2(\vec H\times\vec U)^2+\vec s\cdot\vec U.
\end{equation}
In this paper, we consider the setup where the two external fields are orthogonal to each other, and choose the coordinate system so that they take the constant values
\begin{equation}
\vec H=(0,H,0),\qquad
\vec s=(s,0,0),
\end{equation}
where $H$ and $s$ are the positive moduli of the field vectors. It is then easy to see that the ground state is oriented along the first axis, $\langle\vec U\rangle=(1,0,0)$. We will use the following parameterization that automatically satisfies the constraint on the length of the vector $\vec U(x)$,
\begin{equation}
\vec U(x)=(U^0(x),U^1(x),U^2(x)),\qquad
U^0\equiv\sqrt{1-(U^1)^2-(U^2)^2}.
\end{equation}
A simple manipulation then casts the LO Lagrangian~\eqref{Lagrangian_LO} in the form
\begin{equation}
\mathcal L^{(2)}_\text{eff}=\frac12F^2(\de_\mu\vec U)^2+2F^2HU^2\de_0U^0-\frac12F^2H^2(U^1)^2+sU^0
\label{Lagrangian_LO_exp}
\end{equation}
up to a constant and a surface term. This describes two magnon excitations with the relativistic dispersion relations $\omega_i(\vek p)\equiv\sqrt{\vek p^2+M_i^2}$ and the masses
\begin{equation}
\mii=\sqrt{\frac s{F^2}+H^2},\qquad
\miii=\sqrt{\frac s{F^2}},
\label{masses}
\end{equation}
excited by $U^1$ and $U^2$, respectively. Note that the staggered field makes both magnons massive, in accord with the effect of the quark mass in the chiral perturbation theory~\cite{Gasser:1983yg,Gasser:1984gg}. The magnetic field, on the other hand, only gaps one of the magnons. Moreover, at $s=0$, the gap of this magnon, $\mii=H$, is \emph{exactly} determined by the magnetic field, independently of the microscopic dynamics of the system~\cite{Watanabe:2013uya}.


\section{Setup for evaluation of the free energy}
\label{sec:1loop}

\begin{figure}
\begin{align*}
\text{NLO: }&
\parbox{10mm}{%
\begin{fmfgraph*}(10,10)
\fmfleft{l}
\fmfright{r}
\fmf{plain,right}{l,r,l}
\end{fmfgraph*}}
\,+\,
\parbox{10mm}{%
\begin{fmfgraph*}(10,10)
\fmfleft{l}
\fmfright{r}
\fmf{dashes,right}{l,r,l}
\end{fmfgraph*}}\quad\F1
\qquad\qquad
\parbox{5mm}{%
\begin{fmfgraph*}(5,5)
\fmfleft{l}
\fmfright{r}
\fmf{phantom}{l,v,r}
\fmfdot{v}
\end{fmfgraph*}}\quad\F{\text{c.t.}}\\[2ex]
\text{NNLO: }&
\parbox{20mm}{%
\begin{fmfgraph*}(20,10)
\fmfleft{l}
\fmfright{r}
\fmf{plain,right}{l,v,l}
\fmf{plain,right}{r,v,r}
\fmfdot{v}
\end{fmfgraph*}}\quad\F{2a}
\qquad\qquad
\parbox{20mm}{%
\begin{fmfgraph*}(20,10)
\fmfleft{l}
\fmfright{r}
\fmf{dashes,right}{l,v,l}
\fmf{dashes,right}{r,v,r}
\fmfdot{v}
\end{fmfgraph*}}\quad\F{2b}
\qquad\qquad
\parbox{20mm}{%
\begin{fmfgraph*}(20,10)
\fmfleft{l}
\fmfright{r}
\fmf{plain,right}{l,v,l}
\fmf{dashes,right}{r,v,r}
\fmfdot{v}
\end{fmfgraph*}}\quad\F{2c}\\
&\parbox{30mm}{%
\begin{fmfgraph*}(30,20)
\fmfleft{l}
\fmfright{r}
\fmf{plain,right}{l,v1,l}
\fmf{plain,right}{r,v2,r}
\fmf{dashes,tension=2.5}{v1,v2}
\fmfdot{v1,v2}
\end{fmfgraph*}}\quad\F{2d}
\qquad\qquad
\parbox{15mm}{%
\begin{fmfgraph*}(15,15)
\fmfleft{l}
\fmfright{r}
\fmf{plain,right}{l,r,l}
\fmf{dashes}{l,r}
\fmfdot{l,r}
\end{fmfgraph*}}\quad\F{2e}\\
&\parbox{10mm}{%
\begin{fmfgraph*}(10,10)
\fmftop{t}
\fmfbottom{b}
\fmf{plain,right}{t,b,t}
\fmfv{d.sh=cross,d.si=3mm}{b}
\end{fmfgraph*}}\quad\F{2f}
\qquad\qquad
\parbox{10mm}{%
\begin{fmfgraph*}(10,10)
\fmftop{t}
\fmfbottom{b}
\fmf{dashes,right}{t,b,t}
\fmfv{d.sh=cross,d.si=3mm}{b}
\end{fmfgraph*}}\quad\F{2g}
\qquad\qquad
\parbox{20mm}{%
\begin{fmfgraph*}(20,10)
\fmfleft{l}
\fmfright{r}
\fmf{plain,right}{l,v,l}
\fmf{dashes,tension=2.5}{r,v}
\fmfdot{v}
\fmfv{d.sh=cross,d.si=3mm}{r}
\end{fmfgraph*}}\quad\F{2h}
\end{align*}
\caption{Contributions to the free energy at the NLO and NNLO of the derivative expansion. The dots at line intersections represent interaction vertices, whereas the solid and dashed lines stand for propagators of the modes with mass $\mii$ ($U^1$) and $\miii$ ($U^2$), respectively, see Eq.~\eqref{masses}. The crosses indicate insertion of operators from the NLO Lagrangian~\eqref{Lagrangian_NLO}.}
\label{fig:graphs}
\end{figure}
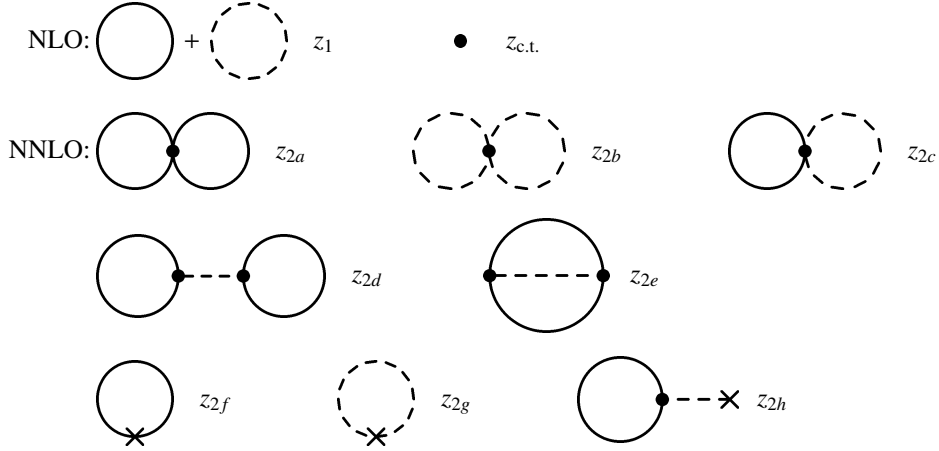

Employing the standard techniques of quantum field theory, the free energy can be most easily evaluated in the Euclidean space using the imaginary time formalism. It then equals minus the sum of all connected vacuum diagrams of the theory~\cite{Kapusta:2006kg}.\footnote{Strictly speaking, the procedure described in the text gives the free energy \emph{density}. We take the liberty to drop the word ``density'' throughout the whole paper as there is no danger of confusing the two closely related quantities.} The contributions to the free energy can, just like the Lagrangian, be organized using the derivative expansion, see Fig.~\ref{fig:graphs}. The LO free energy corresponds to tree-level vacuum diagrams obtained from the LO Lagrangian~\eqref{Lagrangian_LO}. The NLO free energy is given by one-loop diagrams with propagators determined by the LO Lagrangian, and by tree-level diagrams obtained from the NLO Lagrangian~\eqref{Lagrangian_NLO}. Finally, the NNLO free energy contains two-loop diagrams based solely on the LO Lagrangian, one-loop diagrams with an insertion of one operator from the NLO Lagrangian, and NNLO counterterms not shown in Fig.~\ref{fig:graphs}.

\begin{figure}
\fmfset{arrow_len}{2mm}
\begin{equation*}
\begin{alignedat}{3}
\includegraphics[scale=1]{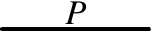}
&=\frac1{F^2}\frac1{P^2+\mii^2}
\qquad\qquad
&
\includegraphics[scale=1]{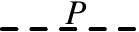}
&=\frac1{F^2}\frac1{P^2+\miii^2}\\
\parbox{15mm}{%
\includegraphics[scale=1]{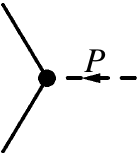}}
&=-2F^2H\omega
\qquad\qquad
&\parbox{15mm}{%
\includegraphics[scale=1]{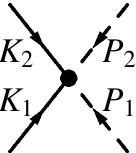}}
&=F^2(K_1+K_2)\cdot(P_1+P_2)-s\\[1ex]
\parbox{15mm}{%
\includegraphics[scale=1]{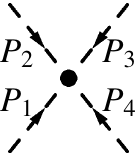}}
&=-F^2(P_1^2+P_2^2+P_3^2+P_4^2)-3s
\qquad\qquad
&\parbox{15mm}{%
\includegraphics[scale=1]{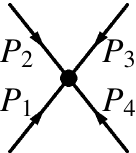}}
&=-F^2(P_1^2+P_2^2+P_3^2+P_4^2)-3s
\end{alignedat}
\end{equation*}
\caption{Feynman rules in Euclidean space following from the LO Lagrangian~\eqref{Lagrangian_LO}. The solid and dashed lines represent $U^1$ and $U^2$, respectively. The Euclidean four-momentum is labeled using uppercase letters whereas the spatial three-momentum is labeled using lowercase letters in bold so that, for example, $P=(\omega,\vek p)$. The arrows indicate the direction of four-momentum.}
\label{fig:FeynRulesLO}
\end{figure}

\begin{figure}
\begin{gather*}
\parbox{5mm}{%
\begin{fmfgraph*}(5,5)
\fmfleft{l}
\fmfright{r}
\fmf{phantom}{l,v,r}
\fmfdot{v}
\end{fmfgraph*}}
=-\left(e_1H^4+e_2H^4+\frac{k_1H^2s}{F^2}+\frac{k_2s^2}{F^4}+\frac{k_3s^2}{F^4}\right)\\
\includegraphics[scale=1]{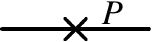}
=a_\text{I}\omega^2+b_\text{I}\vek p^2+c_\text{I}
\qquad\qquad\qquad\qquad
\includegraphics[scale=1]{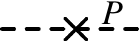}
=a_\text{II}\omega^2+b_\text{II}\vek p^2+c_\text{II}
\end{gather*}
\caption{Feynman rules in Euclidean space following from the NLO Lagrangian~\eqref{Lagrangian_NLO}. The Euclidean four-momentum is labeled using uppercase letters whereas the spatial three-momentum is labeled using lowercase letters in bold so that, for example, $P=(\omega,\vek p)$. The coefficients $a_{\text{I,II}}$, $b_{\text{I,II}}$ and $c_{\text{I,II}}$ are defined in Eq.~\eqref{abc} in terms of the NLO couplings $e_{1,2}$ and $k_{1,2}$.}
\label{fig:FeynRulesNLO}
\end{figure}

For the reader's convenience, we next summarize all the necessary ingredients needed to reproduce our calculation. The nonlinear dependence of the effective Lagrangian on the magnon fields gives rise to an infinite tower of interaction terms. However, only quadratic, cubic and quartic terms in the LO Lagrangian~\eqref{Lagrangian_LO} are needed to determine the free energy up to the NNLO; the corresponding Feynman rules in Euclidean space are given in Fig.~\ref{fig:FeynRulesLO}. Out of the NLO Lagrangian~\eqref{Lagrangian_NLO}, only the constant and quadratic terms are required. The former provide counterterms for the one-loop diagrams contributing to the NLO free energy, whereas the latter enter one-loop diagrams contributing to the NNLO free energy. The required Feynman rules are reviewed in Fig.~\ref{fig:FeynRulesNLO}. For the sake of brevity, we put together all bilinear terms in the NLO Lagrangian that are proportional to squared frequency or momentum, resulting in the following combinations of the NLO effective couplings,
\begin{equation}
\begin{alignedat}{3}
a_\text{I}&\equiv-2\left(2e_1H^2+2e_2H^2+\frac{k_1s}{F^2}\right),
\qquad\qquad
&a_\text{II}&\equiv-2\left(6e_1H^2+6e_2H^2+\frac{k_1s}{F^2}\right),\\
b_\text{I}&\equiv-2\left(2e_1H^2+\frac{k_1s}{F^2}\right),
\qquad\qquad
&b_\text{II}&\equiv-2\left(2e_1H^2+2e_2H^2+\frac{k_1s}{F^2}\right),\\
c_\text{I}&\equiv-2\left(2e_1H^4+2e_2H^4+\frac{3k_1H^2s}{2F^2}+\frac{k_2s^2}{F^4}\right),
\qquad\qquad
&c_\text{II}&\equiv-2\left(\frac{k_1H^2s}{2F^2}+\frac{k_2s^2}{F^4}\right).
\end{alignedat}
\label{abc}
\end{equation}

Naive momentum integration of the loop diagrams leads to ultraviolet divergences that have to be regularized and renormalized. Here we use the standard dimensional regularization in which the spacetime dimension is analytically continued to noninteger values. We adopt the notation $D\equiv4-2\eps$ for the \emph{spacetime} dimension, or equivalently $d\equiv3-2\eps$ for the \emph{spatial} dimension. In the imaginary time formalism, a closed loop gives rise to an integral over spatial momentum and a sum over Matsubara frequencies, which for bosons take the values $\omega_n\equiv 2\pi nT$. We will use the shorthand notation
\begin{equation}
\sumint P\equiv T\sum_{\omega_n}\int_p
\end{equation}
for such a sum-integral. Momentum integrals include an a priori arbitrary renormalization scale $\mu$, which ensures that the integrals have a fixed canonical dimension independent of $\eps$. Thus, the zero-temperature spacetime momentum and spatial momentum integrals are denoted as
\begin{equation}
\int_P\equiv\mu^{2\eps}\int\frac{\dd^D\!\!P}{(2\pi)^D},
\qquad
\int_p\equiv\mu^{2\eps}\int\frac{\dd^d\!\!\vek p}{(2\pi)^d}.
\end{equation}
To renormalize expressions that diverge in the limit $\eps\to0$, we adopt a version of the modified minimal subtraction ($\overline{\text{MS}}$) scheme, in which we do not subtract just the bare poles in $\eps$, but instead re-express simple poles in terms of a particular dimensionless function of $\eps$ that appears in the loop integral $I(M)$, defined below in Eq.~\eqref{Iintegrals},
\begin{equation}
\lambda\equiv\frac{\Gamma(-1+\eps)}{2(4\pi)^{2-\eps}}=-\frac1{32\pi^2}\left[\frac1\eps+1-\gamma_\text{E}+\ln4\pi+\mathcal O(\eps)\right],
\end{equation}
where $\gE\approx0.577$ is the Euler--Mascheroni constant.

Most of the thermal integrals that we shall deal with can be expressed in terms of the basic set of moments of the Bose--Einstein distribution, defined as
\begin{equation}
g_n(M)\equiv\frac1{(4\pi)^2}\left(\frac{4\pi\mu^2}{T^2}\right)^\eps\frac{4\sqrt\pi T^{4-2n}}{\Gamma\bigl(\frac52-n-\eps\bigr)}\int_0^\infty\dd\!x\frac{x^{D-2n}}{\sqrt{x^2+(\beta M)^2}}\frac1{e^{\sqrt{x^2+(\beta M)^2}}-1}.
\label{gndef}
\end{equation}
Likewise, most of the one- and two-loop diagrams (in fact, all diagrams except for the sunset diagram $\F{2e}$) factorize into products of the following two basic one-loop sum-integrals,\footnote{See Ref.~\cite{Andersen:2004fp} for a review of the most frequently occurring sum-integrals.}
\begin{equation}
\begin{split}
I(M)\equiv{}&\sumint P\frac1{P^2+M^2}=\frac1{(4\pi)^2}\left(\frac{4\pi\mu^2}{M^2}\right)^\eps M^2\Gamma(-1+\eps)+g_1(M),\\
\tilde I(M)\equiv{}&\sumint P\frac{\vek p^2}{P^2+M^2}=\frac d2\left[\frac1{(4\pi)^2}\left(\frac{4\pi\mu^2}{M^2}\right)^\eps M^4\Gamma(-2+\eps)+g_0(M)\right].
\end{split}
\label{Iintegrals}
\end{equation}


\subsection{Free energy up to the next-to-leading order}

The LO contribution to the free energy can be directly read off the LO Lagrangian~\eqref{Lagrangian_LO},
\begin{equation}
\boxed{\F{\text{LO}}=\F0=-s-\frac12F^2H^2.}
\label{FLO}
\end{equation}
However, in order to include the effects of nonzero temperature, we have to go to the NLO where loop diagrams start to contribute. Here we have, first of all, the free energy of the free magnon gas, given by
\begin{equation}
\F1=\frac12\sum_{i=\text{I,II}}\sumint P\ln(P^2+M_i^2)=-\frac1d\sum_{i=\text{I,II}}\tilde I(M_i).
\end{equation}
The divergent part of $\F1$ is to be canceled by counterterms from the NLO Lagrangian~\eqref{Lagrangian_NLO}, which read (see Fig.~\ref{fig:FeynRulesNLO})
\begin{equation}
\F{\text{c.t.}}=-\left(e_1H^4+e_2H^4+\frac{k_1H^2s}{F^2}+\frac{k_2s^2}{F^4}+\frac{k_3s^2}{F^4}\right).
\end{equation}
The NLO effective couplings accordingly contain a divergent part, and are renormalized as follows,
\begin{equation}
e_{1,2}=\gamma_{1,2}\Biggl(\lambda+\frac{\bar e_{1,2}}{32\pi^2}\Biggr),\qquad
k_{1,2}=\gamma_{3,4}\Biggl(\lambda+\frac{\bar k_{1,2}}{32\pi^2}\Biggr),\qquad
k_3=\frac{\bar k_3}{32\pi^2}.
\label{renormalization}
\end{equation}
The dimensionless coefficients $\gamma_{1,2,3,4}$ have to be adjusted in order to ensure cancellation of all divergences. The factor $32\pi^2$, on the other hand, is conventional; factoring out this trivial loop factor, we expect the renormalized couplings $\bar e_{1,2}$ and $\bar k_{1,2,3}$ to be of order one. Note that the coupling $k_3$ has no divergent part since the corresponding operator $\vec s^2$ does not depend on the magnon fields and thus is not needed as a counterterm.

Expanding the one-loop free energy $\F1$ in powers of $\eps$, cancellation of divergences in the sum with $\F{\text{c.t.}}$ imposes the following constraints~\cite{Hofmann:2016ucj},
\begin{equation}
\gamma_1+\gamma_2=\frac12,\qquad
\gamma_3=1,\qquad
\gamma_4=1.
\label{gamma}
\end{equation}
The precise values of $\gamma_{1,2}$ will be fixed in the next section by requiring that all temperature-dependent subdivergences in the NNLO free energy get properly subtracted. With all the pieces in place, we can now state the final result for the renormalized free energy at the NLO of the derivative expansion,
\begin{equation}
\boxed{%
\begin{split}
\F{\text{NLO}}={}&\frac{H^4}{64\pi^2}\left(-\frac12+\ln\frac{\mii^2}{\mu^2}+\frac{\bar e_1}3-\frac{4\bar e_2}3\right)+\frac{H^2s}{64\pi^2F^2}\left(-1+2\ln\frac{\mii^2}{\mu^2}-2\bar k_1\right)\\
&+\frac{s^2}{64\pi^2F^4}\left(-1+\ln\frac{\mii^2}{\mu^2}+\ln\frac{\miii^2}{\mu^2}-2\bar k_2-2\bar k_3\right)-\frac12[g_0(\mii)+g_0(\miii)].
\end{split}}
\label{FNLO}
\end{equation}
Although not indicated explicitly in Eq.~\eqref{renormalization}, the renormalized couplings $\bar e_{1,2}$ and $\bar k_{1,2}$ depend on the renormalization scale $\mu$.\footnote{The coupling $\bar k_3$ is not associated with any counterterm and thus, in our renormalization scheme, is scale-independent.} It is easy to check that they satisfy the renormalization group equations $\mu\dd\!\bar e_{1,2}/\!\dd\!\mu=\mu\dd\!\bar k_{1,2}/\!\dd\!\mu=-2$~\cite{Scherer:2002tk}. As a consequence, changing the scale from $\mu_1$ to $\mu_2$ requires to change the couplings according to
\begin{equation}
\bar e_{1,2}(\mu_2)=\bar e_{1,2}(\mu_1)+2\ln\frac{\mu_1}{\mu_2},\qquad
\bar k_{1,2}(\mu_2)=\bar k_{1,2}(\mu_1)+2\ln\frac{\mu_1}{\mu_2}.
\label{RGeq}
\end{equation}
It is a nontrivial consistency check of our calculation that, as a consequence of this running of the LECs, our expression for the NLO free energy~\eqref{FNLO} is independent of the choice of the renormalization scale $\mu$.


\section{Free energy at the next-to-next-to-leading order}
\label{sec:2loop}

The NNLO free energy consists of contributions shown in the last three lines of Fig.~\ref{fig:graphs}. While the sunset diagram $\F{2e}$ is tricky, all the others are straightforward to evaluate, and we therefore just list the results before renormalization,
\begin{align}
\notag
\F{2a}&=\frac1{8F^4}(3s-4F^2\mii^2)[I(\mii)]^2,\\
\notag
\F{2b}&=\frac1{8F^4}(3s-4F^2\miii^2)[I(\miii)]^2,\\
\notag
\F{2c}&=\frac s{4F^4}I(\mii)I(\miii),\\
\label{NNLO}
\F{2d}&=0,\\
\notag
\F{2f}&=\frac1{2F^2}\bigl[(a_\text{I}\mii^2-c_\text{I})I(\mii)+(a_\text{I}-b_\text{I})\tilde I(\mii)\bigr],\\
\notag
\F{2g}&=\frac1{2F^2}\bigl[(a_\text{II}\miii^2-c_\text{II})I(\miii)+(a_\text{II}-b_\text{II})\tilde I(\miii)\bigr],\\
\notag
\F{2h}&=0.
\end{align}
The diagrams $\F{2d}$ and $\F{2h}$ vanish trivially thanks to the fact that the cubic interaction vertex is proportional to frequency carried by the $U^2$ line. The sunset diagram $\F{2e}$ is addressed separately in the next subsection.


\subsection{The sunset diagram}
\label{subsec:sunset}

The evaluation of the sunset diagram $\F{2e}$ represents a nontrivial piece of work, and we therefore give most details needed. Below we present a calculation of the sunset diagram using momentum-space techniques, which allows us to analytically extract the divergent part of the diagram and, at zero temperature, to reduce it to a simple one-dimensional integral that can be easily evaluated numerically. An alternative derivation, utilizing coordinate-space techniques, is described in~\ref{app:sunset}.

To start with, the diagram is given by the following expression in momentum space,
\begin{equation}
\F{2e}=\frac{H^2}{F^2}\,\sumint P\sumint Q\frac{(P_0+Q_0)^2}{(P^2+\mii^2)(Q^2+\mii^2)[(P+Q)^2+\miii^2]}.
\label{sunsetdef}
\end{equation}
Since evaluating the Matsubara sum is the more involved part of the calculation, we remove the frequencies from the numerator by replacing $(P_0+Q_0)^2\to(P+Q)^2+\miii^2-[(\vek p+\vek q)^2+\miii^2]$, which allows us to cast the integral as
\begin{equation}
\F{2e}=\frac{H^2}{F^2}\Bigl\{[I(\mii)]^2-\X(\mii,\miii)\Bigr\},
\qquad
\X(m,M)\equiv\,\sumint P\sumint Q\frac{(\vek p+\vek q)^2+M^2}{(P^2+m^2)(Q^2+m^2)[(P+Q)^2+M^2]}.
\label{defX}
\end{equation}
For the rest, we will follow the method to evaluate massive thermal diagrams put forward in Ref.~\cite{Bugrii:1995vn}. The main trick is to decouple summation over the two Matsubara frequencies in the diagram by using the identity
\begin{equation}
\delta_{P_0+Q_0+K_0}=T\int_0^\beta\dd\!\theta\,e^{\imag\theta(P_0+Q_0+K_0)}.
\end{equation}
This factorizes the integral into three independent Matsubara sums that can be performed easily using the formula
\begin{equation}
T\sum_{\omega_n}\frac{e^{\imag\omega_n\theta}}{\omega_n^2+x^2}=\frac1{2x}\frac{\cosh\Bigl[\Bigl(\frac\beta2-\theta\Bigr)x\Bigr]}{\sinh\frac{\beta x}2}\qquad
\text{for }0\leq\theta\leq\beta.
\end{equation}
Subsequently, the integral over $\theta$ is done, leading to
\begin{equation}
\begin{split}
\X(m,M)={}&\mu^{-2\eps}\int_{p,q,k}(2\pi)^d\delta^d(\vek p+\vek q+\vek k)\frac{\vek k^2+M^2}{4\eps_p\eps_q E_k}\Biggl[\frac1{E_k+\eps_p+\eps_q}\\
&+n(E_k)\left(\frac1{E_k+\eps_p+\eps_q}+\frac1{-E_k+\eps_p+\eps_q}\right)+2n(\eps_p)\left(\frac1{E_k+\eps_p+\eps_q}+\frac1{E_k-\eps_p+\eps_q}\right)\\
&+n(\eps_p)n(\eps_q)\left(\frac1{E_k+\eps_p+\eps_q}+\frac1{E_k-\eps_p+\eps_q}+\frac1{E_k+\eps_p-\eps_q}+\frac1{E_k-\eps_p-\eps_q}\right)\\
&+2n(E_k)n(\eps_p)\left(\frac1{E_k+\eps_p+\eps_q}+\frac1{E_k-\eps_p+\eps_q}+\frac1{-E_k+\eps_p+\eps_q}+\frac1{-E_k-\eps_p+\eps_q}\right)\Biggr],
\end{split}
\label{XmM}
\end{equation}
where we introduced a shorthand notation for the quasiparticle energies and the Bose--Einstein distribution function,
\begin{equation}
\eps_p\equiv\sqrt{\vek p^2+m^2},\qquad
E_p\equiv\sqrt{\vek p^2+M^2},\qquad
n(x)\equiv\frac1{e^{\beta x}-1}.
\end{equation}
The first line in the above expression for $\X(m,M)$ represents the corresponding vacuum diagram. In addition to that, also the two terms with a single Bose factor are divergent, whereas the contributions with two Bose factors are finite. Therefore, only the terms with zero or one Bose factor can depend on the renormalization scale $\mu$.


\subsubsection{Zero Bose factors}
\label{subsec:X0}

\begin{figure}
\begin{center}
\includegraphics[width=0.7\textwidth]{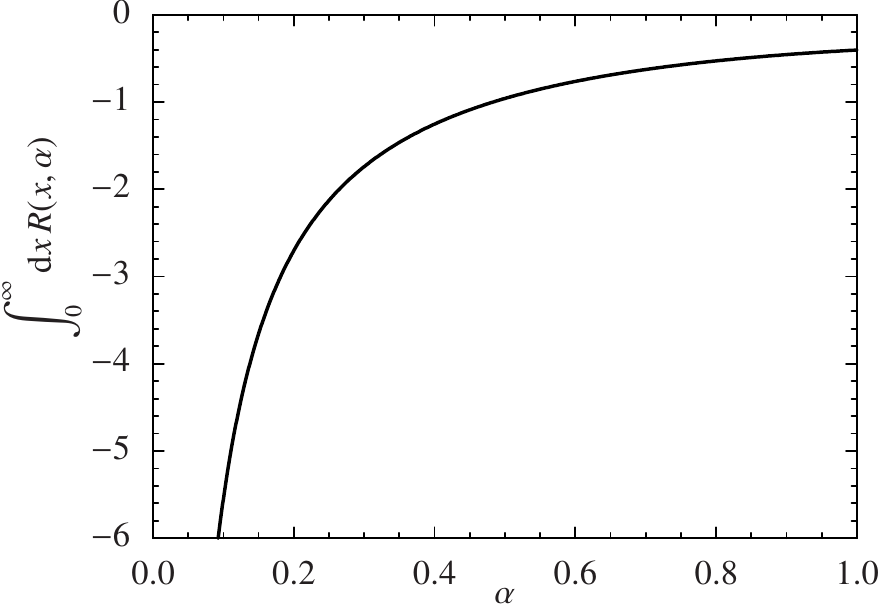}
\end{center}
\caption{Integral of the residual function $R(x,\alpha)$ defined by Eq.~\eqref{Rdef} as a function of the ratio $\alpha=M/m$. Owing to the inequality $\miii\leq\mii$, only the values $0\leq\alpha\leq1$ are physically relevant. Note that the (linear) divergence as $\alpha\to0$ is only fictitious: thanks to the factor $M^2$ in Eq.~\eqref{X0}, the contribution of the integral $I_\text{sun}(m,M)$ to the free energy actually vanishes in the limit $M\to0$.}
\label{fig:Rxalpha}
\end{figure}

The first line of Eq.~\eqref{XmM} can be put into a Lorentz-invariant form by using the identity
\begin{equation}
\frac1{4\eps_p\eps_qE_k}\frac1{E_k+\eps_p+\eps_q}=\int\frac{\dd\!P_0}{2\pi}\frac{\dd\!Q_0}{2\pi}\frac{\dd\!K_0}{2\pi}\frac{2\pi\delta(P_0+Q_0+K_0)}{(P^2+m^2)(Q^2+m^2)(K^2+M^2)}.
\end{equation}
We can now use the full Lorentz (Euclidean) invariance of the zero-temperature part of $\X(m,M)$ to rewrite it as
\begin{equation}
\X_0(m,M)=\frac dD[I_0(m)]^2+\frac1DM^2I_\text{sun}(m,M),
\label{X0}
\end{equation}
where $I_0(m)$ is likewise the zero-temperature part of $I(m)$ and
\begin{equation}
I_\text{sun}(m,M)\equiv\mu^{-2\eps}\int_{P,Q,K}\frac{(2\pi)^D\delta^D(P+Q+K)}{(P^2+m^2)(Q^2+m^2)(K^2+M^2)}
\end{equation}
is the zero-temperature sunset diagram in a theory with nonderivative couplings and different masses. The latter is most conveniently evaluated using coordinate space methods, writing~\cite{Braaten:1995cm}
\begin{equation}
I_\text{sun}(m,M)=\mu^{-2\eps}\int\dd^D\!\!X\,[\Delta(m,X)]^2\Delta(M,X),
\end{equation}
where $\Delta(m,X)$ is the propagator of a free massive particle in the coordinate representation. In $D$ Euclidean dimensions, it can be evaluated explicitly as~\cite{Andersen:2000zn}
\begin{equation}
\Delta(m,X)=\frac{\mu^{2\eps}}{(2\pi)^{2-\eps}}\left(\frac mX\right)^{1-\eps}K_{1-\eps}(mX),
\label{propX}
\end{equation}
where $K_\alpha(x)$ is the modified Bessel function of the second kind. The zero-temperature sunset integral then acquires the dimensionless form
\begin{equation}
I_\text{sun}(m,M)=\left(\frac{\pi\mu^2}{mM}\right)^{2\eps}\frac{m^{1-\eps}M^{1+\eps}}{32\pi^4}\frac{2^{3\eps}}{\Gamma\bigl(\frac D2\bigr)}\int_0^\infty\dd\!x\,x^\eps[K_{1-\eps}(x)]^2K_{1-\eps}\bigl(\tfrac Mmx\bigr).
\label{Isun}
\end{equation}
This is now evaluated following a standard set of steps:
\begin{itemize}
\itemsep0pt
\item Introduce a cutoff $r$ such that $\eps\ll r\ll1$ and split the integral into the ranges $(0,r)$ and $(r,\infty)$.
\item In the integral over $(0,r)$, expand the integrand in powers of $x$ and integrate exactly. Subsequently, expand the result in powers of $\eps$.
\item In the integral over $(r,\infty)$, the limit $\eps\to0$ can be safely taken. Evaluate the resulting integral (see below).
\item Contributions from the two ranges which are singular in the limit $r\to 0$, cancel each other. In the final result, the limit $r\to0$ can therefore be taken.\\[-3ex]
\end{itemize}
As for the integral of $[K_1(x)]^2K_1(\alpha x)$ with $\alpha\equiv M/m$ over $(r,\infty)$, this still does not seem to admit analytic evaluation in a closed form. Since the integral diverges in the limit $r\to0$, it is necessary to extract the divergence first,
\begin{equation}
[K_1(x)]^2K_1(\alpha x)=A(\alpha)\frac1{x^3}+B(\alpha)\frac{\ln x}x+C(\alpha)\frac1x+\mathcal O(x\ln^2x),
\end{equation}
where
\begin{equation}
A(\alpha)=\frac1\alpha,\qquad
B(\alpha)=\frac\alpha2+\frac1\alpha,\qquad
C(\alpha)=\left(-\frac14+\frac\gE2\right)\alpha+\left(-\frac12+\gE-\ln2\right)\frac1\alpha+\frac\alpha2\ln\frac\alpha2.
\end{equation}
Next, we deform the functions multiplying $A(\alpha)$, $B(\alpha)$ and $C(\alpha)$ so as to preserve the divergence structure at $x\to0$ and at the same time to provide an analytically calculable integral converging at $x\to\infty$, for instance
\begin{equation}
[K_1(x)]^2K_1(\alpha x)\equiv A(\alpha)\frac1{x^3}+B(\alpha)\frac{\ln x}xe^{-x^2}+C(\alpha)\frac1{e^x-1}+R(x,\alpha),
\label{Rdef}
\end{equation}
which defines the residuum $R(x,\alpha)$, having an infrared- and ultraviolet-finite integral over the whole range $(0,\infty)$, which has to be evaluated numerically for every given value of $\alpha$ (see Fig.~\ref{fig:Rxalpha} for the numerical values). Putting all the pieces together, the dimensionless zero-temperature sunset integral takes the following form,
\begin{equation}
\begin{split}
\int_0^\infty\dd\!x\,x^\eps[K_{1-\eps}(x)]^2K_{1-\eps}(\alpha x)={}&-\frac1{16\eps^2}\left(\alpha+\frac2\alpha\right)+\frac1{16\eps}\left[\left(3\alpha-\frac2\alpha\right)\ln\alpha+(-2+\gE-\ln2)\left(\alpha+\frac2\alpha\right)\right]\\
&+\frac1{16}\Biggl\{-\frac12\left(\alpha+\frac2\alpha\right)\ln^2\alpha+\left[(2+5\gE-5\ln2)\alpha+(-4+2\gE-2\ln2)\frac1\alpha\right]\ln\alpha\\
&+\left(-2-2\gE+\frac{9\gE^2}2-\frac{\pi^2}{12}+2\ln2-7\gE\ln2+\frac72\ln^22\right)\left(\alpha+\frac2\alpha\right)\Biggr\}\\
&+\int_0^\infty\dd\!x\,R(x,\alpha)+\mathcal O(\eps).
\end{split}
\label{Kaux}
\end{equation}

\subsubsection{One Bose factor}

The second line of Eq.~\eqref{XmM} can be simplified by using the identities
\begin{equation}
\begin{split}
\frac1{4\eps_p\eps_q}\left(\frac1{E_k+\eps_p+\eps_q}+\frac1{-E_k+\eps_p+\eps_q}\right)&=\int\frac{\dd\!P_0}{2\pi}\frac{\dd\!Q_0}{2\pi}\frac{2\pi\delta(P_0+Q_0+K_0)}{(P^2+m^2)(Q^2+m^2)}\Biggr|_{K^2=-M^2},\\
\frac1{4\eps_qE_k}\left(\frac1{E_k+\eps_p+\eps_q}+\frac1{E_k-\eps_p+\eps_q}\right)&=\int\frac{\dd\!Q_0}{2\pi}\frac{\dd\!K_0}{2\pi}\frac{2\pi\delta(P_0+Q_0+K_0)}{(Q^2+m^2)(K^2+M^2)}\Biggr|_{P^2=-m^2}.
\end{split}
\end{equation}
The part of $\X(m,M)$ containing a single Bose factor then takes the form
\begin{equation}
\X_1(m,M)=\int_k(\vek k^2+M^2)\frac{n(E_k)}{E_k}\int_P\frac1{(P^2+m^2)[(P+K)^2+m^2]}\Biggr|_{K^2=-M^2}+\int_p\frac{2n(\eps_p)}{\eps_p}\int_K\frac{\vek k^2+M^2}{(K^2+M^2)[(K+P)^2+m^2]}\Biggr|_{P^2=-m^2}.
\end{equation}
The inner one-loop integrals can be evaluated using the standard Feynman parameterization. The final result is somewhat lengthy, but straightforward to obtain,
\begin{equation}
\begin{split}
\X_1(m,M)={}&\left(\frac{4\pi\mu^2}{T^2}\right)^\eps\frac{\Gamma(\eps)}{(4\pi)^2}\Biggl\{\left[M^2g_1(M)+\frac d2g_0(M)\right][1+\eps\K1(\beta m,\beta M)]\\
&+g_1(m)\left[-m^2+\frac12M^2+2\eps T^2\K2(\beta m,\beta M)\right]+dg_0(m)\left[\frac13+\eps\K3(\beta m,\beta M)\right]\Biggr\}+\mathcal O(\eps),
\end{split}
\label{X1}
\end{equation}
where
\begin{equation}
\begin{split}
\K1(a,b)&\equiv-\int_0^1\dd\!x\ln\bigl[a^2-b^2x(1-x)\bigr],\\
\K2(a,b)&\equiv\frac12\int_0^1\dd\!x\,\Bigl\{\bigl[3a^2x^2+b^2(1-3x)\bigr]\ln\bigl[a^2x^2+b^2(1-x)\bigr]-\bigl[a^2x^2+b^2(1-x)\bigr]\Bigr\},\\
\K3(a,b)&\equiv-\int_0^1\dd\!x\,x^2\ln\bigl[a^2x^2+b^2(1-x)\bigr].
\end{split}
\label{Kndef}
\end{equation}


\subsubsection{Two Bose factors}

The last two lines of Eq.~\eqref{XmM} constitute the only piece of the two-loop free energy that cannot be reduced to elementary one-dimensional integrals and has to be evaluated numerically. The computing effort required is equivalent to that of evaluating a three-dimensional integral,
\begin{equation}
\begin{split}
\X_2(m,M)={}&\int\frac{\dd^3\!\!\vek p}{(2\pi)^3}\frac{\dd^3\!\!\vek q}{(2\pi)^3}\frac{\vek k^2+M^2}{4\eps_p\eps_q E_k}\Biggl[n(\eps_p)n(\eps_q)\Biggl(\frac1{E_k+\eps_p+\eps_q}+\frac1{E_k-\eps_p+\eps_q}+\frac1{E_k+\eps_p-\eps_q}+\frac1{E_k-\eps_p-\eps_q}\Biggr)\\
&+2n(E_k)n(\eps_p)\Biggl(\frac1{E_k+\eps_p+\eps_q}+\frac1{E_k-\eps_p+\eps_q}+\frac1{-E_k+\eps_p+\eps_q}+\frac1{-E_k-\eps_p+\eps_q}\Biggr)\Biggr],
\end{split}
\label{X2}
\end{equation}
where $\vek k=-(\vek p+\vek q)$. The whole sunset diagram $\F{2e}$ is then determined by Eqs.~\eqref{defX}, \eqref{X0}, \eqref{Isun}, \eqref{Kaux}, \eqref{X1} and \eqref{X2} together with
\begin{equation}
\X(m,M)=\X_0(m,M)+\X_1(m,M)+\X_2(m,M).
\end{equation}


\subsection{Renormalized free energy at NNLO}

The full free energy at NNLO is given by a sum of the pieces listed in Eq.~\eqref{NNLO} and our above result for the sunset diagram. It is a nontrivial check of consistency that all temperature-dependent subdivergences contributing to the free energy at NNLO cancel for the values of the counterterms $\gamma_{3,4}$ shown in Eq.~\eqref{gamma}, if we in addition set~\cite{Hofmann:2016ucj}
\begin{equation}
\gamma_1=-\frac16,\qquad
\gamma_2=\frac23.
\end{equation}
The full NNLO free energy then takes the form
\begin{equation}
\F{\text{NNLO}}=\F{\text{NNLO,ren}}+\F{\text{NNLO,div}},
\end{equation}
where
\begin{equation}
\begin{split}
\F{\text{NNLO,div}}={}&\frac{H^6}{F^2}\left[\lambda^2+\frac\lambda{32\pi^2}\left(-\frac16+2\bar e_2\right)\right]+\frac{H^4s}{F^4}\left[\frac{3\lambda^2}2+\frac\lambda{32\pi^2}\left(-\frac{11}6+\frac{2\bar e_1}3+\frac{4\bar e_2}3+\bar k_1\right)\right]\\
&+\frac{H^2s^2}{F^6}\left[-\frac{\lambda^2}2+\frac\lambda{32\pi^2}\left(-\frac{23}{12}+\frac{5\bar e_1}3-\frac{14\bar e_2}3+2\bar k_2\right)\right]+\frac{s^3}{F^8}\left[\frac\lambda{32\pi^2}\bigl(-4\bar k_1+4\bar k_2\bigr)\right].
\end{split}
\end{equation}
These remaining local divergences have to be canceled by adding counterterms from the next-to-next-to-leading-order Lagrangian. We do not attempt to even classify all operators contributing to this order-six Lagrangian, since the mere number of operators is known to be about a hundred in the case of chiral perturbation theory~\cite{Fearing:1994ga,Bijnens:1999sh}. However, since the divergent part of the free energy is a polynomial in the external fields, it is obvious that the divergences \emph{can} be canceled by a suitable set of counterterms consistent with all the symmetries of the system.

The finite, renormalized part of the NNLO free energy, $\F{\text{NNLO,ren}}$, is given by a lengthy expression, and we therefore display it piece by piece. First of all, it is useful to distinguish the zero-temperature and nonzero-temperature contributions,
\begin{equation}
\F{\text{NNLO,ren}}=\F{\text{NNLO,ren,0}}+\F{\text{NNLO,ren,T}},
\end{equation}
where
\begin{equation}
\boxed{%
\begin{split}
\F{\text{NNLO,ren,0}}&=\frac1{1024\pi^4}\left(c_{6,0}\frac{H^6}{F^2}+c_{4,1}\frac{H^4s}{F^4}+c_{2,2}\frac{H^2s^2}{F^6}+c_{0,3}\frac{s^3}{F^8}\right)-\frac{H^2}{128\pi^4F^2}\mii\miii^3\int_0^\infty\dd\!x\,R(x,\miii/\mii),\\
\F{\text{NNLO,ren,T}}&=\frac1{32\pi^2}\left(d_{4,0}\frac{H^4}{F^2}+d_{2,1}\frac{H^2s}{F^4}+d_{0,2}\frac{s^2}{F^6}\right)+\frac1{32\pi^2}\left(e_{2,0}\frac{H^2}{F^2}+e_{0,1}\frac{s}{F^4}\right).
\end{split}}
\label{FNNLO}
\end{equation}
The individual coefficients of the expansion are given by
\begin{align}
\notag
c_{6,0}={}&\frac14-\frac\gE6+\frac16\ln4\pi-\frac23\ln\frac{\mii^2}{\mu^2}-\ln^2\frac{\mii^2}{\mu^2}+\bar e_2\left(\frac13+2\ln\frac{\mii^2}{\mu^2}\right),\\
\notag
c_{4,1}={}&\frac{11}4+\frac{13\gE}6-5\gE^2-\frac{\pi^2}6+\frac{11}6\ln4\pi-4\ln2+8\gE\ln2-4\ln^22-\frac{13}3\ln\frac{\mii^2}{\mu^2}-\frac32\ln^2\frac{\mii^2}{\mu^2}\\
\notag
&+\left(\frac{2\bar e_1}3+\bar k_1\right)\ln\frac{\mii^2}{\mu^2}+\frac{2\bar e_2}3\left(1+2\ln\frac{\mii^2}{\mu^2}\right),\\
\notag
c_{2,2}={}&\frac{23}8+\frac{49\gE}{12}-\frac{15\gE^2}2-\frac{\pi^2}4+\frac{23}{12}\ln4\pi-6\ln2+12\gE\ln2-6\ln^22\\
\notag
&-\frac{14}3\ln\frac{\mii^2}{\mu^2}+2(\gE-\ln2)\ln\frac{\mii^2}{\miii^2}-\frac32\ln^2\frac{\mii^2}{\mu^2}+2\ln\frac{\mii^2}{\mu^2}\ln\frac{\miii^2}{\mu^2}\\
\notag
&+\bar e_1\left(-\frac16+\frac23\ln\frac{\mii^2}{\mu^2}+\ln\frac{\miii^2}{\mu^2}\right)+\bar e_2\left(1-\frac23\ln\frac{\mii^2}{\mu^2}-4\ln\frac{\miii^2}{\mu^2}\right)-\bar k_1\ln\frac{\mii^2}{\miii^2}+2\bar k_2\ln\frac{\mii^2}{\mu^2},\\
c_{0,3}={}&-\frac12\ln^2\frac{\mii^2}{\miii^2}-2(\bar k_1-\bar k_2)\left(\ln\frac{\mii^2}{\mu^2}+\ln\frac{\miii^2}{\mu^2}\right),\\
\notag
d_{4,0}={}&2g_1(\mii)\left(-1+\ln\frac{\mii^2}{T^2}\right),\\
\notag
d_{2,1}={}&g_1(\mii)\left(-1+\frac32\ln\frac{\mii^2}{\mu^2}-\ln\frac{T^2}{\mu^2}+\frac{\bar e_1}3-\frac{4\bar e_2}3+\frac{\bar k_1}2\right)\\
\notag
&+g_1(\miii)\left[2+\frac12\ln\frac{\mii^2}{\mu^2}+2\ln\frac{T^2}{\mu^2}+\bar e_1-4\bar e_2+\frac{\bar k_1}2-2\K1(\beta\mii,\beta\miii)\right],\\
\notag
d_{0,2}={}&g_1(\mii)\left(-\frac12\ln\frac{\mii^2}{\miii^2}-\bar k_1+\bar k_2\right)+g_1(\miii)\left(\frac12\ln\frac{\mii^2}{\miii^2}-\bar k_1+\bar k_2\right),\\
\notag
e_{2,0}={}&2g_0(\mii)\left[1+\ln\frac{T^2}{\mu^2}-\bar e_2-3\K3(\beta\mii,\beta\miii)\right]+g_0(\miii)\left[3+3\ln\frac{T^2}{\mu^2}+\bar e_1-4\bar e_2-3\K1(\beta\mii,\beta\miii)\right]\\
\notag
&+16\pi^2g_1^2(\mii)-4T^2g_1(\mii)\K2(\beta\mii,\beta\miii)-32\pi^2\X_2(\mii,\miii),\\
\notag
e_{0,1}={}&-4\pi^2\left[g_1(\mii)-g_1(\miii)\right]^2.
\end{align}
Just like in the case of the NLO free energy~\eqref{FNLO}, it is a nontrivial consistency check that the temperature-dependent part of the renormalized NNLO free energy, $\F{\text{NNLO,ren,T}}$, is independent of the renormalization scale $\mu$ by virtue of the flow equations~\eqref{RGeq}. The zero-temperature part $\F{\text{NNLO,ren,0}}$ contains a residual $\mu$-dependence, which is expected to be eliminated by the corresponding $\mu$-dependence of the NNLO effective couplings, not included here.

\begin{figure}
\begin{center}
\includegraphics[width=0.7\textwidth]{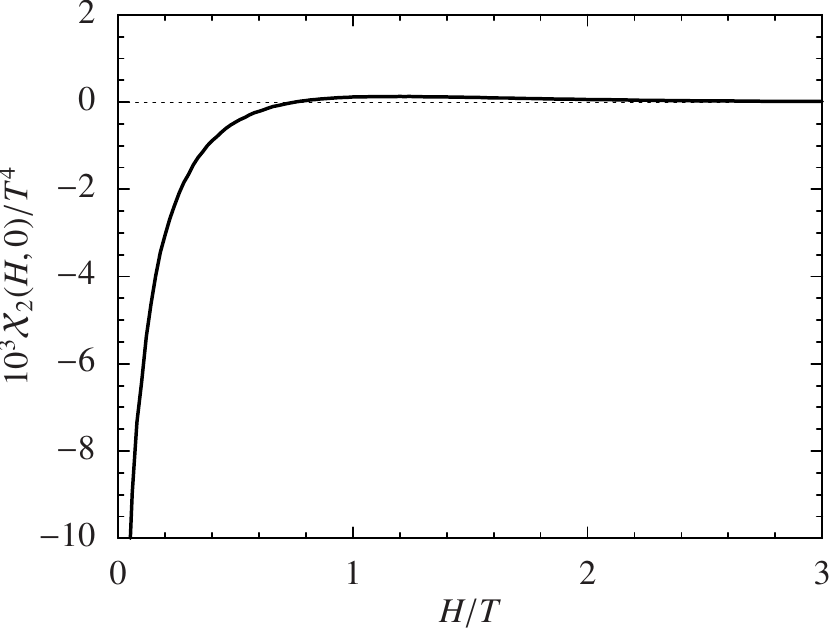}
\end{center}
\caption{The thermal integral $\X_2$, defined by Eq.~\eqref{X2}, as a function of the magnetic field at $s=0$. Both $\X_2$ and $H$ are made dimensionless by rescaling by an appropriate power of temperature, and $\X_2$ is in addition rescaled by a factor of 1000 to get more natural units on the vertical axis. The value of $\X_2$ drops quickly with increasing magnetic field, and for $H\gtrsim T$ it makes a negligible contribution to the coefficient $\tilde e_{2,0}$.}
\label{fig:X2s0}
\end{figure}

Given how lengthy the full result for the NNLO free energy is, it may be of interest to spell out explicitly some special cases that allow for more tractable expressions. The case of $H=0$ is particularly simple since the sunset diagram then becomes zero, and the NNLO renormalized free energy is given by
\begin{equation}
\F{\text{NNLO,ren}}\Bigr|_{H=0}=-(\bar k_1-\bar k_2)\left[\frac{s^3}{256\pi^4F^8}\ln\frac s{F^2\mu^2}+\frac{s^2}{16\pi^2F^6}g_1(\sqrt s/F)\right].
\end{equation}
However, the staggered field may not be easy to implement as a tunable external field in experiment; we rather expect it to be a fixed parameter of a given antiferromagnetic material. Assuming absence of spin-orbit coupling or other perturbations that would break the $\gr{SO}(3)$ symmetry explicitly, it makes sense to set $s=0$ and focus on the dependence on the magnetic field $H$. The renormalized NNLO free energy then becomes
\begin{equation}
\F{\text{NNLO,ren}}\Bigr|_{s=0}=\frac{H^6}{1024\pi^4F^2}\tilde c_{6,0}+\frac{H^2}{32\pi^2F^2}\tilde e_{2,0},
\end{equation}
where
\begin{equation}
\begin{split}
\tilde c_{6,0}&=\frac14-\frac\gE6+\frac16\ln4\pi-\frac23\ln\frac{H^2}{\mu^2}-\ln^2\frac{H^2}{\mu^2}+\bar e_2\left(\frac13+2\ln\frac{H^2}{\mu^2}\right),\\
\tilde e_{2,0}&=2g_0(H)\left(\frac13+\ln\frac{H^2}{\mu^2}-\bar e_2\right)+\frac{\pi^2T^4}{45}\left(3+3\ln\frac{H^2}{\mu^2}+\bar e_1-4\bar e_2\right)+16\pi^2g_1^2(H)-32\pi^2\X_2(H,0).
\end{split}
\label{FNNLOs0}
\end{equation}
The numerical values of the function $\X_2(H,0)$ are displayed in Fig.~\ref{fig:X2s0}; the other thermal factors in Eq.~\eqref{FNNLOs0} are given by simple one-dimensional integrals that are trivial to evaluate numerically. Without doing so explicitly, we just remark that the $\X_2$ term in Eq.~\eqref{FNNLOs0} is negligible compared to the other contributions to $\tilde e_{2,0}$ for $H\gtrsim T$.

\section{Summary and conclusions}
\label{sec:summary}

In this paper, we have utilized EFT to compute the free energy of three-dimensional antiferromagnets in crossed magnetic and staggered fields at the two-loop order. The individual LO, NLO and NNLO contributions to the free energy are given by Eqs.~\eqref{FLO}, \eqref{FNLO} and \eqref{FNNLO}, respectively. Renormalization of the free energy has been carried out analytically, and the result thus depends on a set of renormalized NLO couplings, denoted as $\bar e_{1,2}$ and $\bar k_{1,2,3}$, in addition to the sole LO effective coupling $F$, corresponding to the square root of the spin stiffness. The zero-temperature part of the free energy is accordingly given by a set of closed analytic expressions except for a single one-dimensional integral of the function $R(x,\alpha)$, defined in Eq.~\eqref{Rdef}, whose numerical values are shown in Fig.~\ref{fig:Rxalpha}. The final result for the free energy has been shown to be independent of the arbitrary renormalization scale $\mu$.

The thermal part of the free energy takes a semi-analytic form. First, it depends on a series of one-dimensional integrals, $g_n$ defined by Eq.~\eqref{gndef} and $\K n$ defined by Eq.~\eqref{Kndef}, which are straightforward to compute numerically and we thus leave their detailed evaluation up to the reader. Second, there is a single piece of the sunset diagram that cannot be reduced to a simple one-dimensional integral and is encoded in the function $\X_2$, defined by Eq.~\eqref{X2}. This function corresponds to a three-dimensional integral in momentum space, which is suppressed by two Bose factors and thus is exponentially convergent. It has to be evaluated using a suitable numerical integrator; the values of this function along the section $s=0$ are shown in Fig.~\ref{fig:X2s0}.

The zero-temperature part of the NNLO result in Eq.~\eqref{FNNLO} in principle has to be augmented with a set of local counterterms, stemming from the NNLO effective Lagrangian. Given the expected high number of operators contributing to the NNLO Lagrangian, this reduces somewhat the predictive power of our EFT at the NNLO accuracy. However, we would like to emphasize that the logarithmic terms in the NNLO free energy~\eqref{FNNLO}, as well as its whole thermal part, are free from such an ambiguity and thus represent a genuine prediction of our calculation.

Of course, having evaluated the free energy of the system of interest is just the first step in the analysis of its thermodynamic properties. The free energy can in turn be used to generate many other thermodynamical observables in equilibrium such as the staggered magnetization (the order parameter for symmetry breaking) and magnetization (the response of the antiferromagnet to the external magnetic field). The analysis of these observables building upon the result for the free energy obtained here will be the subject of a companion paper~\cite{companion}.


\appendix

\section{Construction of the effective Lagrangian}
\label{app:lagrangian}

In this appendix, we shall justify the expressions for the LO and NLO Lagrangians given in Eqs.~\eqref{Lagrangian_LO} and~\eqref{Lagrangian_NLO}. We divide the argument into two steps. First we discuss how the symmetries of the underlying microscopic dynamics are reflected in the low-energy effective theory. In the second step, we then classify all operators that contribute to the effective Lagrangian at the LO and NLO of the derivative expansion.


\subsection{Symmetries of the effective action}

The dynamics of spin systems possesses an internal global $\gr{SO}(3)$ symmetry in the absence of spin-orbit coupling and other perturbations. While this alone would be sufficient to fix the dependence of the effective action on the magnon field $\vec U(x)$, the background magnetic and staggered fields break the symmetry. We therefore have to constrain the way that the EFT depends on these background fields.

We will follow the philosophy of Refs.~\cite{Leutwyler:1993gf,Leutwyler:1993iq}. The response of the microscopic dynamics to the background fields $\vec H(x)$ and $\vec s(x)$ can be described by a generating functional $\Gamma[\vec H,\vec s]$, which collects all connected Green's functions of the local operators that these fields couple to. The low-energy observables of the system can equally well be described by a low-energy EFT, which can contain completely different dynamical degrees of freedom than the microscopic theory, but is coupled to the \emph{same} background fields. In our case, the low-energy degrees of freedom are contained in the magnon field $\vec U(x)$, and the effective action is then a local functional, $S_\text{eff}[\vec U,\vec H,\vec s]$. Although the dynamical degrees of freedom in the microscopic theory and in the low-energy EFT are different, their generating functionals have to coincide, that is,
\begin{equation}
e^{\imag\Gamma[\vec H,\vec s]}=\frac 1Z\int\mathcal D\vec U\,e^{\imag S_\text{eff}[\vec U,\vec H,\vec s]}.
\end{equation}
The general logic in constructing the EFT therefore is that the symmetries of the microscopic dynamics imply certain symmetries of the generating functional $\Gamma[\vec H,\vec s]$, which in turn constrain the form of the effective action $S_\text{eff}[\vec U,\vec H,\vec s]$.

Our task therefore is to characterize the symmetries of the generating functional $\Gamma[\vec H,\vec s]$. To that end, we write down a generic microscopic model for a spin system, following Ref.~\cite{fradkin2013field}. The dynamics of a \emph{single} spin in presence of an external magnetic field $\vec H$ can be described by a coherent state path integral with the classical action
\begin{equation}
S[\vec n,\vec H]=JS_\text{WZ}[\vec n]+J\int\dd\!t\,\vec H\cdot\vec n(t),
\end{equation}
where $J$ is the magnitude of the particle's spin, $\vec n(t)$ a unit vector variable, and $S_\text{WZ}$ is the Wess--Zumino (WZ) action, encoding the Berry phase of the spin state in the external field~\cite{Altland:book}. It is straightforward to generalize the action to an arbitrary collection of spins placed at fixed positions labeled by the coordinate $\vek x$,
\begin{equation}
S[\{\vec n\},\vec H]=J\sum_{\vek x}S_\text{WZ}[\vec n(\vek x)]+J\sum_{\vek x}\int\dd\!t\,\vec H\cdot\vec n(\vek x,t)-gJ^2\sum_{(\vek x,\vek x')}\int\dd\!t\,\vec n(\vek x,t)\cdot\vec n(\vek x',t),
\end{equation}
where $g$ is the spin--spin coupling. Finally, for spins localized to bipartite lattices it is possible to introduce a staggered field $\vec s$, coupling to the alternating (staggered) sum of the spins. We will also allow for both external fields to depend on time in an arbitrary way, thus writing the action as
\begin{equation}
S[\{\vec n\},\vec H,\vec s]=J\sum_{\vek x}S_\text{WZ}[\vec n(\vek x)]+J\sum_{\vek x}\int\dd\!t\,\bigl[\vec H(t)+(-1)^{\vek x}\vec s(t)\bigr]\cdot\vec n(\vek x,t)-gJ^2\sum_{(\vek x,\vek x')}\int\dd\!t\,\vec n(\vek x,t)\cdot\vec n(\vek x',t).
\label{Smicro}
\end{equation}
It is obvious that this action is invariant under simultaneous global (time-independent) rotations of the spins and both external fields. However, a much stronger statement actually holds. It is known that the simple $\vec H\cdot\vec n$ coupling added to the WZ action makes the action invariant under simultaneous \emph{gauge} (time-dependent) rotations of the spin and the $\vec H$-field~\cite{Frohlich:1993gs,Bar:2004bw}. A simple way to understand this is to note that the magnetic field couples to the total spin, which is a conserved charge of the $\gr{SO}(3)$ symmetry~\cite{Leutwyler:1993gf}. We therefore conclude that the action~\eqref{Smicro} is invariant under simultaneous gauge transformations of the spin field and \emph{both} external fields, provided that the latter transform as
\begin{equation}
\delta\vec H(t)=\vec\eps(t)\times\vec H(t)+\de_0\vec\eps(t),\qquad
\delta\vec s(t)=\vec\eps(t)\times\vec s(t),
\label{genfuncsym}
\end{equation}
where $\vec\eps(t)$ is the infinitesimal parameter of the time-dependent rotation. Since the spin field $\vec n(\vek x,t)$ is integrated over in the path integral, Eq.~\eqref{genfuncsym} defines the desired symmetry of the generating functional $\Gamma[\vec H,\vec s]$. The action of the low-energy EFT should thus be invariant under simultaneous gauge transformations of the magnon field $\vec U$ and the external fields $\vec H$ and $\vec s$; here $\vec H$ plays the role of a temporal gauge field of the $\gr{SO}(3)$ symmetry and $\vec s$ behaves as a covariant vector field.

Note that the nonlocal spin--spin interaction in the microscopic action~\eqref{Smicro} does not allow for a further extension of the symmetry to \emph{coordinate}-dependent gauge transformations. However, this can be achieved upon taking the continuum limit, where the spin--spin coupling becomes simply the spatial part of the kinetic term of the magnons.


\subsection{Classification of operators at the leading and next-to-leading order}

With the above in mind, we will now construct the most general effective action with the following field content:
\begin{itemize}
\itemsep0pt
\item$\vec U(x)$, the magnon field.
\item$\vec A_\mu(x)\equiv(\vec H(x),\vec{\vek 0})$, the gauge field of $\gr{SO}(3)$.
\item$\vec s(x)$, the external vector field.
\end{itemize}
We demand that the action has $\gr{SO}(3)$ gauge invariance. In addition, we impose Poincar\'e invariance, that is, Lorentz invariance augmented with spacetime translation invariance. For the sake of simplicity, we also assume that spatial parity is preserved.

The effective action is organized according to the number of derivatives acting on the magnon fields. For consistency with the assumed gauge invariance, the gauge field $\vec A_\mu$ counts as one derivative. Also, the staggered field $\vec s$ counts as two derivatives since it turns out to be proportional to the \emph{squared} mass of the magnons. Using the unit vector field $\vec U$ is, however, not the best way to go about the classification of operators contributing to the effective Lagrangian. The reason is that $\vec U$ itself does not contain any derivative, and thus in principle operators with an arbitrarily high number of $\vec U$ factors can contribute at any fixed order in the derivative expansion. At the same time, since magnons are Nambu--Goldstone bosons, their interactions have to contain derivatives. It is therefore more practical to use field variables that make this manifest.

Following Ref.~\cite{Andersen:2014ywa}, we first map the vector $\vec U$ to a $2\times2$ matrix variable $U$ via $\vec U\cdot\vec\sigma=U\sigma_1U^{-1}$, where $\vec\sigma$ is the set of Pauli matrices. Subsequently, we introduce the variable $\phi_\mu$ via
\begin{equation}
\phi_\mu\equiv\frac\imag2U^{-1}(D_\mu U^2)U^{-1}=\frac\imag2\bigl[(D_\mu U)U^{-1}+U^{-1}(D_\mu U)\bigr],
\end{equation}
where $D_\mu$ is the $\gr{SO}(3)$-covariant derivative, constructed using the gauge field $\vec A_\mu$. The field $\phi_\mu\equiv\phi_\mu^a\sigma_a$ with $a=2,3$ plays the role of the covariant derivative of the magnon, and transforms in the vector representation of the unbroken subgroup $\gr{SO}(2)$, generated by $\sigma_1$. Likewise, the field-strength tensor of the $\gr{SO}(3)$ gauge field, $\vec F_{\mu\nu}\equiv\de_\mu\vec A_\nu-\de_\nu\vec A_\mu+\vec A_\mu\times\vec A_\nu$, is mapped on a matrix field $F_{\mu\nu}\equiv\vec F_{\mu\nu}\cdot\vec\sigma$ and subsequently traded for the tensor field $G_{\mu\nu}$ defined by
\begin{equation}
G_{\mu\nu}\equiv U^{-1}F_{\mu\nu}U+\imag[\phi_\mu,\phi_\nu]-D_\mu\phi_\nu+D_\nu\phi_\mu.
\end{equation}
This field, $G_{\mu\nu}\equiv G^\alpha_{\mu\nu}\sigma_\alpha$ with $\alpha=1$, plays the role of a field-strength tensor of the unbroken subgroup $\gr{SO}(2)$, and thus transforms as a singlet thereof. Finally, the staggered field $\vec s$ can be mapped on the matrix field
\begin{equation}
\Xi=\Xi^i\sigma_i\equiv U^{-1}(\vec s\cdot\vec\sigma)U,
\end{equation}
where $i=1,2,3$. This field transforms as a direct sum of a singlet and a vector under the unbroken subgroup $\gr{SO}(2)$.

It has been shown that assuming Lorentz invariance and the absence of anomalies, invariance of the effective action under a symmetry group automatically implies invariance of the corresponding effective Lagrangian in three spatial dimensions~\cite{Leutwyler:1993iq}. The invariant effective Lagrangian can then be constructed as a polynomial in $\phi^a_\mu$, $G^\alpha_{\mu\nu}$, $\Xi^i$ and their covariant derivatives~\cite{Andersen:2014ywa}. In the derivative counting, $\phi^a_\mu$ is of order one whereas $G^\alpha_{\mu\nu}$ and $\Xi^i$ are of order two, hence only a finite number of operators contributes to the Lagrangian at any fixed order in the derivative expansion. The precise form of the operators is constrained by Lorentz invariance and by the \emph{unbroken} subgroup $\gr{SO}(2)$.


\subsubsection{Leading-order Lagrangian}

At the leading, second-order of the derivative expansion, Lorentz invariance restricts the possible operators in the Lagrangian to
\begin{equation}
\phi^a_\mu\phi^{b\mu},\qquad
\xcancel{D_\mu\phi^{a\mu}},\qquad
\Xi^i.
\end{equation}
The crossed operator is a total derivative and thus can be dropped, whereas the remaining two operators, once projected onto singlets of the unbroken subgroup $\gr{SO}(2)$, are equivalent to $(D_\mu\vec U)^2$ and $\vec s\cdot\vec U$. These are the operators constituting the LO Lagrangian~\eqref{Lagrangian_LO}.


\subsubsection{Next-to-leading order Lagrangian}

At the next-to-leading order of the derivative expansion, the following operators are allowed by Lorentz invariance, modulo surface terms and redundancy due to integration by parts~\cite{Andersen:2014ywa},
\begin{equation}
\begin{gathered}
\phi^a_\mu\phi^{b\mu}\phi^c_\nu\phi^{d\nu},\qquad
\xcancel{\epsilon^{\kappa\lambda\mu\nu}\phi^a_\kappa\phi^b_\lambda\phi^c_\mu\phi^d_\nu},\qquad
\underline{\phi^{a\mu}\phi^{b\nu}D_\mu\phi^c_\nu},\qquad
\xcancel{\epsilon^{\kappa\lambda\mu\nu}\phi^a_\kappa\phi^b_\lambda D_\mu\phi^c_\nu},\qquad
D_\mu\phi^a_\nu D^\mu\phi^{b\nu},\qquad
D_\mu\phi^{a\mu}D_\nu\phi^{b\nu},\\
\phi^{a\mu}\phi^{b\nu}G^\alpha_{\mu\nu},\qquad
\xcancel{\epsilon^{\kappa\lambda\mu\nu}\phi^a_\kappa\phi^b_\lambda G^\alpha_{\mu\nu}},\qquad
\underline{D^\mu\phi^{a\nu}G^\alpha_{\mu\nu}},\qquad
G^\alpha_{\mu\nu}G^{\beta\mu\nu},\qquad
\Xi^i\Xi^j,\qquad
\Xi^i\phi^a_\mu\phi^{b\mu},\qquad
\Xi^iD_\mu\phi^{a\mu}.
\end{gathered}
\end{equation}
The crossed operators are odd under parity and are thus ruled out by our assumption of parity conservation. In addition, the underlined operators are obviously in contradiction with the  unbroken $\gr{SO}(2)$ invariance. The remaining operators have to be translated back into the physical field variables $\vec U$, $\vec H$ and $\vec s$. We will not show all details here as the argument closely parallels that given in Ref.~\cite{Andersen:2014ywa}, but merely list the correspondence between independent operators in the two notations,
\begin{equation}
\begin{split}
\phi^a_\mu\phi^{b\mu}\phi^c_\nu\phi^{d\nu}&\to(D_\mu\vec U\cdot D^\mu\vec U)^2\quad\text{and}\quad(D_\mu\vec U\cdot D_\nu\vec U)^2,\\
D_\mu\phi^a_\nu D^\mu\phi^{b\nu},D_\mu\phi^{a\mu}D_\nu\phi^{b\nu}&\to D_\mu D^\mu\vec U\cdot D_\nu D^\nu\vec U\quad\text{and}\quad\vec F_{\mu\nu}\cdot\vec F^{\mu\nu},\\
\phi^{a\mu}\phi^{b\nu}G^\alpha_{\mu\nu},G^\alpha_{\mu\nu}G^{\beta\mu\nu}&\to\vec F_{\mu\nu}\cdot(D^\mu\vec U\times D^\nu\vec U)\quad\text{and}\quad(\vec F_{\mu\nu}\cdot\vec U)^2,\\
\Xi^i\Xi^j&\to\vec s^2\quad\text{and}\quad(\vec s\cdot\vec U)^2,\\
\Xi^i\phi^a_\mu\phi^{b\mu}&\to(\vec s\cdot\vec U)(D_\mu\vec U)^2,\\
\Xi^iD_\mu\phi^{a\mu}&\to(\vec s\times\vec U)\cdot(D_\mu D^\mu\vec U).
\end{split}
\end{equation}
The list of linearly independent operators that can appear in the NLO Lagrangian therefore reads
\begin{equation}
\begin{gathered}
(D_\mu\vec U\cdot D^\mu\vec U)^2,\qquad
(D_\mu\vec U\cdot D_\nu\vec U)^2,\qquad
\xcancel{D_\mu D^\mu\vec U\cdot D_\nu D^\nu\vec U},\qquad
\underline{\vec F_{\mu\nu}\cdot\vec F^{\mu\nu}},\qquad
\underline{\vec F_{\mu\nu}\cdot(D^\mu\vec U\times D^\nu\vec U)},\\
\underline{(\vec F_{\mu\nu}\cdot\vec U)^2},\qquad
\vec s^2,\qquad
(\vec s\cdot\vec U)^2,\qquad
(\vec s\cdot\vec U)(D_\mu\vec U)^2,\qquad
\xcancel{(\vec s\times\vec U)\cdot(D_\mu D^\mu\vec U)}.
\end{gathered}
\end{equation}
However, the crossed operators become redundant with the others upon using the equation of motion, following from the LO Lagrangian~\eqref{Lagrangian_LO}. In addition, recalling the definition of the background gauge field $\vec A_\mu$, it is obvious that $\vec F_{\mu\nu}=0$, which disposes of the underlined operators. This leaves us with altogether five operators that contribute to the NLO Lagrangian for our system, as shown in Eq.~\eqref{Lagrangian_NLO}. Note that at this order, there are no independent operators (and thus no unknown coupling constants) required by the presence of the external magnetic field; the dependence on this field is fully determined by the structure of the covariant derivatives.


\section{Coordinate space evaluation of the sunset diagram}
\label{app:sunset}

In this appendix, we outline an alternative evaluation of the sunset diagram $\F{2e}$, following a method developed in Ref.~\cite{Gerber:1988tt}. In the coordinate space, the sunset diagram can be represented as
\begin{equation}
\F{2e}=\frac{2H^2}{F^2}\int_{\cal T}\dd^D\!\!X\,\Iup G(X)\de_0\Iup G(X)\de_0\IIup G(X),
\label{z6D}
\end{equation}
cf.~Eq.~\eqref{sunsetdef}, where $G(X)$ stands for the thermal propagator of a free massive relativistic particle and the superscript $\Iup{}$ or $\IIup{}$ indicates which of the two magnon modes the propagator refers to. The integral extends over the torus ${\cal T}\equiv S^1\times\mathbb{R}^d$ with the circle $S^1$ defined by $-\beta/2\leq X_0\leq\beta/2$. The first step of the analysis is to decompose the propagator into the zero-temperature part $\Delta$, given in Eq.~\eqref{propX}, and the thermal part, denoted as $\overline G$,
\begin{equation}
G(X)=\Delta(X)+\overline G(X).
\label{finiteTzeroT}
\end{equation}
Substituting this decomposition into Eq.~\eqref{z6D} converts the integral therein to
\begin{equation}
\begin{split}
\int_{\cal T}\dd^D\!\!X\,\Bigl(&\Iup{\overline G}\de_0\Iup{\overline G}\de_0\IIup{\overline G}+\Iup\Delta\de_0\Iup{\overline G}\de_0\IIup{\overline G}+\Iup{\overline G}\de_0\Iup\Delta\de_0\IIup{\overline G}+\Iup{\overline G}\de_0\Iup{\overline G}\de_0\IIup\Delta\\
&+\Iup\Delta\de_0\Iup{\overline G}\de_0\IIup\Delta+\Iup\Delta\de_0\Iup\Delta\de_0\IIup{\overline G}+\Iup{\overline G}\de_0\Iup\Delta\de_0\IIup\Delta+\Iup\Delta\de_0\Iup\Delta\de_0\IIup\Delta\Bigl),
\end{split}
\label{sunsetDecomp}
\end{equation}
and the resulting terms are next processed one by one. The four integrals on the first line are ultraviolet-convergent. However, the integrals on the second line are divergent in the limit $D\to4$. In order to isolate the divergences, we cut out a sphere $\cal S$ of radius smaller than $\beta/2$ around the origin of the $X$-space, that is, further decompose the integrals as
\begin{equation}
\int_{\cal T}\dd^D\!\!X\to\int_{\cal S}\dd^D\!\!X+\int_{\cal T\setminus S}\dd^D\!\!X.
\end{equation}
The integrals over the complement $\cal T\setminus S$ are well-defined in the limit $D\to4$. In the integrals over the sphere $\cal S$, we then perform a series of subtractions which allows us to concentrate the divergences in a set of auxiliary integrals that can be evaluated analytically. To that end, we subtract the first few terms of the Taylor expansion of the \emph{thermal} propagators around the origin $X=0$. Thus, in the first two integrals on the second line of Eq.~\eqref{sunsetDecomp}, containing $\de_0\overline G$, subtraction of the first nontrivial term of the Taylor series is sufficient,
\begin{equation}
\de_0\overline G\to\de_0\overline G-X_0\de_0^2\overline G(0).
\end{equation}
In the third integral, containing $\overline G$ without a derivative, the first two nontrivial terms of the Taylor series have to be subtracted,
\begin{equation}
\overline G\to\overline G-g_1-\frac12X^\mu X^\nu\de_\mu\de_\nu\overline G(0).
\end{equation}
Noting further that the second derivatives of the thermal propagator at the origin can be rewritten in terms of the basic thermal integrals $g_n$, defined in Eq.~\eqref{gndef}, as\footnote{See Eq.~(3.5) in Ref.~\cite{Gerber:1988tt}.}
\begin{equation}
\de_\mu\de_\nu\overline G(0)=-\frac12\delta_{\mu\nu}g_0+\delta_{\mu0}\delta_{\nu0}\left(\frac D2g_0+M^2g_1\right),
\end{equation}
the integrals over the sphere $\cal S$ take the form
\begin{equation}
\begin{split}
\int_{\cal S}\dd^D\!\!X\,\Iup\Delta\de_0\Iup{\overline G}\de_0\IIup\Delta={}&\int_{\cal S}\dd^D\!\!X\,\Iup\Delta\left[\de_0\Iup{\overline G}-X_0\left(\frac32g_0^\text{I}+\mii^2g_1^\text{I}\right)\right]\de_0\IIup\Delta+\int_{\cal S}\dd^D\!\!X\,\Iup\Delta X_0\left(\frac32g_0^\text{I}+\mii^2g_1^\text{I}\right)\de_0\IIup\Delta,\\
\int_{\cal S}\dd^D\!\!X\,\Iup\Delta\de_0\Iup\Delta\de_0\IIup{\overline G}={}&\int_{\cal S}\dd^D\!\!X\,\Iup\Delta\de_0\Iup\Delta\left[\de_0\IIup{\overline G}-X_0\left(\frac32g_0^\text{II}+\miii^2g_1^\text{II}\right)\right]+\int_{\cal S}\dd^D\!\!X\,\Iup\Delta\de_0\Iup\Delta X_0\left(\frac32g_0^\text{II}+\miii^2g_1^\text{II}\right),\\
\int_{\cal S}\dd^D\!\!X\,\Iup{\overline G}\de_0\Iup\Delta\de_0\IIup\Delta={}&\int_{\cal S}\dd^D\!\!X\,\left[\Iup{\overline G}-g_1^\text{I}+\frac14(\vek x^2-3X_0^2)g_0^\text{I}-\frac12X_0^2\mii^2g_1^\text{I}\right]\de_0\Iup\Delta\de_0\IIup\Delta\\
&+\int_{\cal S}\dd^D\!\!X\,\left[g_1^\text{I}-\frac14(\vek x^2-3X_0^2)g_0^\text{I}+\frac12X_0^2\mii^2g_1^\text{I}\right]\de_0\Iup\Delta\de_0\IIup\Delta.
\end{split}
\end{equation}
The respective first integrals on the right-hand side, in which the subtraction has been performed, are now convergent in the limit $D\to4$. In order to reduce the respective second terms on the right-hand side to analytically calculable integrals, we finally extend the integration domain to the whole Euclidean space $\mathbb{R}^D$ by rewriting
\begin{equation}
\int_{\cal S}\dd^D\!\!X\to\int\dd^D\!\!X-\int_{\mathbb{R}^D\setminus\cal S}\dd^D\!\!X.
\end{equation}
The integrals over $\mathbb{R}^D\setminus\cal S$ are again convergent in the limit $D\to4$. All ultraviolet divergences are thereby contained in a set of zero-temperature Euclidean integrals that can be evaluated analytically,
\begin{equation}
\begin{split}
R_1&\equiv\int\dd^D\!\!X\,X_0\Iup\Delta(X)\de_0\IIup\Delta(X),\\
R_2&\equiv\int\dd^D\!\!X\,X_0\Iup\Delta(X)\de_0\Iup\Delta(X),\\
R_3&\equiv\int\dd^D\!\!X\,\de_0\Iup\Delta(X)\de_0\IIup\Delta(X),\\
R_4&\equiv\int\dd^D\!\!X\,X_0^2\de_0\Iup\Delta(X)\de_0\IIup\Delta(X),\\
R_5&\equiv\int\dd^D\!\!X\,\vek x^2\de_0\Iup\Delta(X)\de_0\IIup\Delta(X).\\
\end{split}
\end{equation}
The last piece in Eq.~\eqref{sunsetDecomp} involves three zero-temperature propagators. Following the same logic as above, it can be decomposed as
\begin{equation}
\int_{\cal T}\dd^D\!\!X\,\Iup\Delta\de_0\Iup\Delta\de_0\IIup\Delta=\int\dd^D\!\!X\,\Iup\Delta\de_0\Iup\Delta\de_0\IIup\Delta-\int_{\mathbb{R}^D\setminus\cal T}\dd^D\!\!X\,\Iup\Delta\de_0\Iup\Delta\de_0\IIup\Delta.
\end{equation}
The first piece corresponds to the zero-temperature sunset integral, evaluated explicitly in Sec.~\ref{subsec:X0} of the main text, and will be denoted simply as $C$ below.

Collecting all the various pieces, the final expression for the sunset diagram reads
\begin{equation}
\begin{split}
\F{2e}={}&\frac{2H^2}{F^2}\left(\int_{\cal T}\dd^4\!\!X\,T+\int_{\cal T\setminus S}\dd^4\!\!X\,U+\int_{\cal S}\dd^4\!\!X\,V-\int_{\mathbb{R}^D\setminus\cal S}\dd^4\!\!X\,W\right)+\frac{2H^2}{F^2}(R+C),\\
T\equiv{}&\Iup{\overline G}\de_0\Iup{\overline G}\de_0\IIup{\overline G}+\Iup\Delta\de_0\Iup{\overline G}\de_0\IIup{\overline G}+\Iup{\overline G}\de_0\Iup\Delta\de_0\IIup{\overline G}+\Iup{\overline G}\de_0\Iup{\overline G}\de_0\IIup\Delta,\\
U\equiv{}&\Iup\Delta\de_0\Iup{\overline G}\de_0\IIup\Delta+\Iup\Delta\de_0\Iup\Delta\de_0\IIup{\overline G}+\Iup{\overline G}\de_0\Iup\Delta\de_0\IIup\Delta+\Iup\Delta\de_0\Iup\Delta\de_0\IIup\Delta,\\
V\equiv{}&\Iup\Delta\left[\de_0\Iup{\overline G}-X_0\left(\frac32g_0^\text{I}+\mii^2g_1^\text{I}\right)\right]\de_0\IIup\Delta+\Iup\Delta\de_0\Iup\Delta\left[\de_0\IIup{\overline G}-X_0\left(\frac32g_0^\text{II}+\miii^2g_1^\text{II}\right)\right]\\
&+\left[\Iup{\overline G}-g_1^\text{I}+\frac14(\vek x^2-3X_0^2)g_0^\text{I}-\frac12X_0^2\mii^2g_1^\text{I}\right]\de_0\Iup\Delta\de_0\IIup\Delta,\\
W\equiv{}&\Iup\Delta X_0\left(\frac32g_0^\text{I}+\mii^2g_1^\text{I}\right)\de_0\IIup\Delta+\Iup\Delta\de_0\Iup\Delta X_0\left(\frac32g_0^\text{II}+\miii^2g_1^\text{II}\right)+\left[g_1^\text{I}-\frac14(\vek x^2-3X_0^2)g_0^\text{I}+\frac12X_0^2\mii^2g_1^\text{I}\right]\de_0\Iup\Delta\de_0\IIup\Delta\\
&+\Iup\Delta\de_0\Iup\Delta\de_0\IIup\Delta,\\
R\equiv{}&\left(\frac32R_1+\frac34R_4-\frac14R_5\right)g_0^\text{I}+\frac32R_2g_0^\text{II}+\left(R_1\mii^2+R_3+\frac12R_4\mii^2\right)g_1^\text{I}+\miii^2R_2g_1^\text{II}.
\end{split}
\end{equation}
The only pieces of the expression for $\F{2e}$ that are ultraviolet-divergent are $R$ and $C$, which can be evaluated analytically. The rest is finite in the limit $D\to4$ and can be computed numerically for given values of the external fields and the low-energy coupling $F$. A nontrivial check of the numerical evaluation is that the result cannot depend on the size of the sphere $\cal S$.


\bibliographystyle{elsarticle-num} 
\bibliography{references}

\end{fmffile}
\end{document}